# Artistic control over the glitch in AI-generated motion capture

Jamal Knight
Animal Logic Academy
University of Technology Sydney
Sydney NSW Australia
jamal.knight@student.uts.edu.au

Andrew Johnston
Animal Logic Academy
University of Technology Sydney
Sydney NSW Australia
andrew.johnston@uts.edu.au

Adam Berry
Data Science Institute
University of Technology Sydney
Sydney NSW Australia
adam.berry@uts.edu.au

## ABSTRACT

Artificial intelligence (AI) models are prevalent today and provide a valuable tool for artists. However, a lesser-known artifact that comes with AI models that is not always discussed is the glitch. Glitches occur for various reasons; sometimes, they are known, and sometimes they are a mystery. Artists who use AI models to generate art might not understand the reason for the glitch but often want to experiment and explore novel ways of augmenting the output of the glitch. This paper discusses some of the questions artists have when leveraging the glitch in AI art production. It explores the unexpected positive outcomes produced by glitches in the specific context of motion capture and performance art.

## CCS CONCEPTS

• Human-centered computing • Human-centered computing~Human computer interaction (HCI) • Human-centered computing~Human computer interaction (HCI)~interaction paradigms • Human-centered computing~Human computer interaction (HCI)~Interaction paradigms~Collaborative interaction

## KEYWORDS

Motion capture, machine learning, AI, performance art, keypoint detection, animation.

## 1   Introduction

Motion capture is used in many industries, but is perhaps most known for the entertainment industries. The movement of the apes in 'Dawn of the Planet of the Apes', Neytiri in 'Avatar' or the Hulk in 'The Avengers' were all driven by motion capture technology. The motion of performers is accurately calculated by a myriad of cameras surrounding a capture stage. This style of art needs to be accurate to the millimetre, or the performance is at risk of falling into the 'uncanny valley' of animated motion leading to the audience becoming disengaged.

There are other styles of art whose representation of animation does not require accuracy to this detail, however. Performance art is one domain where lower cost, faster-setup, reduced-accuracy motion capture, for instance, has found a home, where approximated outputs are sufficient to drive (often abstract) animations of dancer movement, suitable for projection or integration into the performance space.

An emerging technology in this domain is AI-powered human pose detection, which can identify coarse skeletal movement using only a small number of consumer-grade RGB cameras – an output of sufficient quality for building a fully animated mesh that reflects human movement [7]. This research explores the application of machine learning in this area, using the single-camera VIBE [8] and multi-camera EasyMocap [5] models to drive performer-driven abstract animation.  While conducting this research, there has been a realisation of the value and power of embracing the glitches such systems deliver.  Glitches are valued by artists; they are historically important and motivate artists to experiment [1] [2] [3] [6] [10]. But understanding the form and function of those glitches, what drives them, and how to replicate them, is critical to enabling artists to harness these unpredictable surprises in an artistic environment.

While we emphasize the artistic value of glitches in AI-generated motion capture, we acknowledge the need to address the lack of transparency and understanding when glitches occur due to the black box nature of AI models. To better support artists in their creative process, we propose the exploration of explainable features that would shed light on the occurrence and nature of glitches, enabling performers to have a deeper understanding of these phenomena. By incorporating XAI techniques, we can empower artists to have more control over glitches, allowing them to intentionally exploit and manipulate unexpected outcomes to achieve their desired artistic goals. In the context of this workshop, we speculate on the potential of XAI to enhance the creative exploration of glitches in AI art production and foster a deeper connection between artists and the underlying AI models.

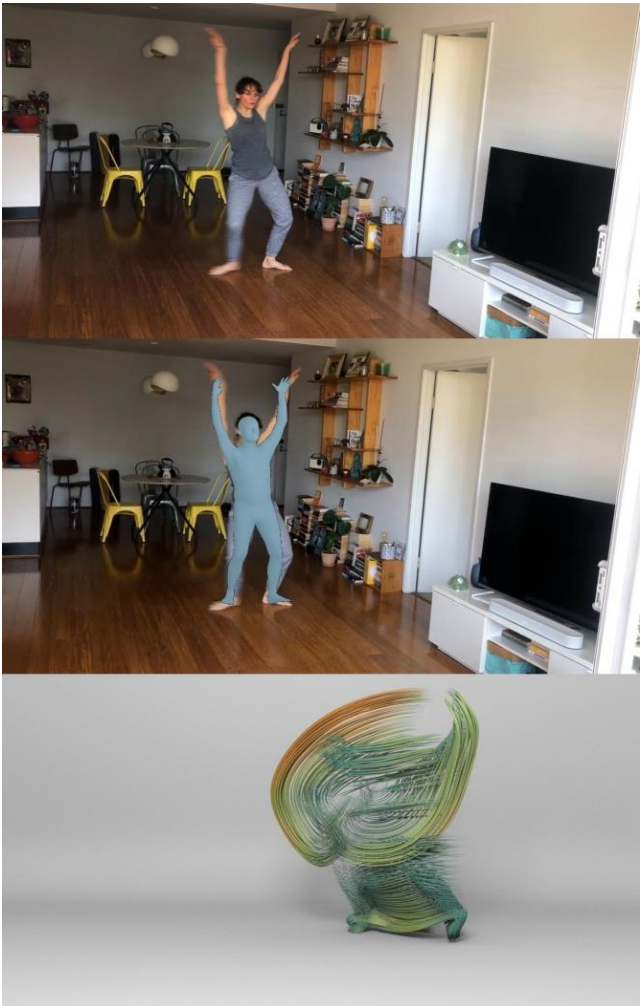

**Figure 1: Visual representations of key stages in the VIBE model, the raw video input of the subject dancing (top), the mesh generation of the VIBE model superimposed over the input video before temporal smoothness is applied (centre), and a frame of the abstract animation applied to the output mesh (lower).**

## 2 Errors vs the glitch

The glitch can be described as '... that which creates minor disturbances without actually damaging its major functioning. Glitches do not stop transmission: they merely make it scrappy, dirty or noisy' [4]. In the book Glitch Art in Theory and Practice, [2] mentions that glitch art originated before 1939, ranging from generative music and sculpture to modern digital art. [3] says, 'Today's digital technology enables artists to explore new territories for content by capturing and examining the area beyond the boundary of "normal" functions and uses of software.'

It is essential to distinguish between when a model fails to run and when the output is unexpected. Unexpected outputs in AI-generated art can be a delightful surprise by bringing new meaning to a piece or revealing an interesting and unexpected aspect. These rare occurrences could be an error in the code, a quirk of the AI model or a user error by the artist. Sometimes there is no way to know where the glitch comes from besides further experiments and re-runs. The original authors of the model have probably encountered the same unexpected results, but rarely are these documented or publicised (perhaps because they are likely viewed as errors to be suppressed or corrected).  To artists, however, these accidental errors and strange outputs can sometimes be the most exciting output the model can produce.

## 3 Examples of the glitch in AI motion capture

### 3.1 Controlling the ghost

In this paper, we describe creative projects which result in motion capture driven animations that are displayed as a performer dances. The machine learning models that are used output an animated mesh, which is editable. The keyframes can be smoothed, and the animation can be fine-tuned after the mesh is generated. This is ideal for artists as they have some degree of control of the mesh before abstract animation is applied.

When tracking performer movements to create animated meshes for performance art, interesting glitches have emerged on occasion:

If four cameras capture a subject's motion, and the subject walks out of view of two or three cameras, the animated mesh will behave in a way that floats around the screen.
The animated mesh can glitch and behave strangely if the camera calibration step is miscalculated. The camera calibration step in detecting human pose is important because it enables calculation of where the physical cameras are in space. If one or more of the cameras are miscalculated, the animated mesh floats and dances around the frame uncontrollably.

The above can result in an ethereal embodiment of a floating ghost-like motion. It is as if the motion capture subject is haunted by their animated ghost floating nearby. The animated figure will occasionally respond if the subject rotated suddenly or changed direction. The animation felt like it was trying to match the subject's movement for a second before giving up and continuing on its floaty path. A video example is available here [A].

Noting the potential value in this context, an attempt was made to control the number and type of glitches by purposefully adding incorrect data for the camera calibration. This was essentially an informed trial-and-error process, where the impacts of varying calibration parameters on the movement of the ghostly mesh we observed. The more incorrect calibration data was used for the cameras, the more detached the mesh became from the subject. A video example of the glitches merged with the correct pose detection output is available here [B].

Without deep knowledge of the underlying AI model, the ability to control and manipulate glitches to achieve artistic outcomes is constrained to exactly this type of trial-and-error methodology. If insight into the causes and drivers of the glitch were more readily available and their relationship to controllable parameters known, then it would both open up the types of artistic experimentation available and increase the efficiency of generating useful outputs. In the context of pose detection glitches, exaggerating or reducing unexpected movement and the ability to balance the mix between expected and unexpected outputs would prove particularly powerful and enable greater artistic control.

### 3.2 Testing the ghost

Feedback from choreographers gathered during collaboration indicates that they have been captivated by the glitches that AI pose detection models produce. Running Machine [9], an Australian and Japanese co-production produced by Sam McGilp, Harrison Hall, Yuiko Masukawa, Makoto Uemura and Kazuhiko Hiwa, featured the glitchy nature of AI-generated pose detection. The producers appreciated the glitch to the extent that they would try to force or exaggerate the effect. Two subjects would be recorded on video in front of a green screen, with one subject completely covered in green, who would move the other subject in various positions. This also 'confused' the AI model into producing the glitch effect, sometimes detecting the greenscreen subject and sometimes detecting the other subject. Other experiments were conducted in front of a green screen, with the top half of one subject in green and the bottom half of the other in green. A similar effect was produced. The choreographers requested an animated mesh of the glitched result without any smoothing or cleanup. A rendering of the mesh was projected on a screen as part of the performance. Examples of these glitches are available here [C] [D].

### 3.3 Confusing the ghost

An example where glitch artifacts were produced unexpectedly was when AI-based motion capture attempted to capture a performer on slings. The slings were attached to beams on the ceiling, and the performers would swing gracefully in different formations. The expected result would be a near-accurate representation of the performers swinging through the air. However, the AI model produced a very glitchy result. The reason for the glitch became apparent when the performer dismounted the slings when the animated mesh snapped back to the performer. It was as if the model could not understand the performer in the slings but automatically recognised them when they were walking on solid ground. Upon inspection, it was revealed that the data the model was trained on was extensive but did not include performers in slings or similar situations. This new information for the model caused it to glitch. An example of this is available here [E]. An artistic representation of glitch animation is available here [F] An entire playlist of all the glitches is available here [G].

## 4 What artists need

No AI model is perfect, and flaws are discovered after some testing. Authors of AI models primarily show examples of their models working seamlessly. However, for artists, it would be helpful to show (with examples) where the model will glitch or behave unexpectantly. In layman's terms, an accompanying explanation describing the reason for the anomaly would be helpful to avoid or exploit the unexpected result. This information would be valuable when choosing an AI model from the outset.

If Github repositories were more honest and forthcoming with the various outcomes of their models, artists would be more experimental with them. It's great that a pose detection model can produce an animated mesh of a human walking, but can it produce motion that a human cannot? Which parameters can I adjust to 'break' the model and cause it to fly around the space like a ragdoll? The ability to unlock these secret abilities in the model is akin to using cheat codes in a video game to access different ways to navigate the game. It may not be what the author intended, but it is often creatively productive to experiment with.

The area where there is no control is the model itself and the glitches it produces. It is unknown whether the environmental, performance or system parameters to tune which will help shape the types of interesting glitches that might power the art. This relatively untapped source of creativity holds considerable potential for innovative experimentation. Although they are not the intended outcome, glitch artifacts should be embraced, and AI practitioners are encouraged to provide methods where this phenomenon occurs.

## ACKNOWLEDGMENTS

Animal Logic Academy, Cloe Fournier, Box of Birds, Entagma.com and Carlos Barreto. This research is supported by an Australian Government Research Training Program Scholarship.